\documentclass[sigconf,10pt]{acmart}

\usepackage{balance}
\usepackage{float}
\usepackage{algorithm}
\usepackage{algpseudocode}  
\usepackage{hyperref}
\usepackage{url}
\usepackage{graphicx}
\usepackage{subcaption}
\usepackage{amsmath}
\usepackage{etoolbox}
\usepackage{tikz}
\usepackage{lipsum}
\setcopyright{cc}
\setcctype{by-nd}
\copyrightyear{2025}
\acmYear{2025}
\acmDOI{10.1145/3714394.3754435}
\acmConference[UbiComp Companion '25]{Companion of the 2025 ACM International Joint Conference on Pervasive and Ubiquitous Computing}{October 12--16, 2025}{Espoo, Finland}
\acmBooktitle{Companion of the 2025 ACM International Joint Conference on Pervasive and Ubiquitous Computing, October 12--16, 2025, Espoo, Finland}
\acmISBN{979-8-4007-1477-1/2025/10}

\begin{document}

\title{SocialPulse: An On-Smartwatch System for Detecting Real-World Social Interactions}

\author{Md Sabbir Ahmed, Arafat Rahman, Mark Rucker, Laura E. Barnes}
\email{{msabbir, jgh6ds, mr2an, lb3dp}@virginia.edu}
\affiliation{%
  \department{Department of Systems and Information Engineering}
  \institution{University of Virginia}
  \city{Charlottesville}
  \state{VA}
  \country{USA}
}

\keywords{Watch, On-Device System, Social Interaction, Audio Process}



\begin{abstract}

\noindent Social interactions are a fundamental part of daily life and play a critical role in well-being. As emerging technologies offer opportunities to unobtrusively monitor behavior, there is growing interest in using them to better understand social experiences. However, automatically detecting interactions—particularly via wearable devices—remains underexplored. Existing systems are often limited to controlled environments, constrained to in-person interactions, and rely on rigid assumptions such as the presence of two speakers within a fixed time window. These limitations reduce their generalizability to capture diverse real-world interactions. To address these challenges, we developed a real-time, on-watch system capable of detecting both in-person and virtual interactions. The system leverages transfer learning to detect foreground speech (FS) and infers interaction boundaries based upon FS and conversational cues like whispering. In a real-world evaluation involving 11 participants over a total of 38 days (Mean = 3.45 days, SD = 2.73), the system achieved an interaction detection accuracy of 73.18\%. Follow-up with six participants indicated perfect recall for detecting interactions. These preliminary findings demonstrate the potential of our system to capture interactions in daily life—providing a foundation for applications such as personalized interventions targeting social anxiety.
\end{abstract}

\maketitle

\section{Introduction}
 
\noindent Social interaction plays a key role in well-being and has been shown to support mental health, for example, by reducing stress \cite{Liang2024-nt, Ono2011-ep}. In contrast, negative social interactions can adversely affect health \cite{Shahrestani2015-dm, Schwerdtfeger2009-jy}. An individual’s mental health status, including momentary changes like state anxiety during interactions, can fluctuate \cite{Wang_Larrazabal_Rucker_Barnes_2023, Larrazabal_Wang_Rucker_2025}. Similarly, patterns of interaction may also vary depending on an individual's mental health status \cite{Dissing2019-pd}. Therefore, understanding real-world social interaction patterns is essential for both advancing mental health research and enabling personalized interventions that promote mental well-being in everyday life.


One method for capturing information about a person's social interactions and their patterns (e.g., sequences of speech and silence) can be through self-reported data. However, such data can be biased and burdensome due to their subjective and effortful nature \cite{Lucas2021-bk, Roos2023-wk}. A promising alternative is passive sensing, which can collect real-time data unobtrusively and with minimal user burden. Several studies have used smartphones to passively detect social interactions \cite{Lane_2011, Katevas2019-xv}. Yet, these efforts are limited by small sample sizes (e.g., N=5 in \cite{Lane_2011}) and evaluations in controlled environments \cite{Katevas2019-xv}, raising concerns about generalizability. Moreover, smartphone-based systems perform lower compared to smartwatch-based approaches in detecting interactions, as shown in a recent study \cite{Liang2023-gi}. A key reason may be that smartphones are not always close to the body - for example, they are often stored in pockets - resulting in lower quality acoustic data \cite{Liang2023-gi}. In contrast, smartwatches are more consistently worn in different contexts and locations \cite{Shahmohammadi2017-hm}, probably due to their comfort and convenience, which makes them suitable for continuous interaction detection.

Despite growing interest in social sensing, there are still very few studies focused on real-time social interaction detection using smartwatches, and existing efforts face several limitations. For example, a promising system by \cite{Liang2023-gi} is restricted to detecting only in-person interactions, even though virtual interactions have become a significant—and for some, preferred—part of daily life \cite{Hutchins_Allen_Curran_Kannis-Dymand_2021}. Additionally, their system assumes that at least two people will speak within every 30-second window during an interaction, an assumption that may not hold true always in real-world settings. Other work, such as \cite{Boateng2019-gt}, is constrained by very limited data—only 3.5 hours of smartwatch wear—raising concerns about the system’s ability to generalize across varied environments and timescales (e.g., interactions at home during the evening). Meanwhile, studies like \cite{Rahman2011-kx, Bari2018-os} rely on physiological signals, such as respiration rate, to detect conversations. However, these approaches face practical limitations in real-world deployment. Commercial smartwatches, such as the Samsung Galaxy Watch, do not include respiration sensors \cite{Wikipedia_contributors2025-qs}. While some devices like Fitbit and Apple Watch estimate respiration rate, it is typically measured during sleep \cite{noauthor-fitbit-breathing-rate, noauthor-undated-av-apple}, when social interaction does not occur. Furthermore, these systems require hardware like chest-worn belts \cite{Rahman2011-kx, Bari2018-os}, which needs to be worn in specific positions. These setups can be obtrusive and uncomfortable—especially over extended periods or under tight clothing—making them less feasible for long-term, everyday use.

To address the limitations of existing systems, we introduce SocialPulse, an on-smartwatch system that detects social interactions in real time by processing short segments of acoustic data using a duty-cycled approach. In a real-world deployment with 11 participants, SocialPulse achieved an interaction detection accuracy of 73.18\%, with no false negatives reported by six participants during follow-up feedback. This work advances ubiquitous computing in the following ways:

\begin{itemize}
    \item To our knowledge, this is the first on-watch system capable of detecting both in-person and virtual social interactions, enabling the capture of a broader range of everyday conversations and offering deeper insights into individuals’ social patterns.
    
    \item Instead of relying on the number of speakers within each time window, our system uses conversational cues (e.g., laughter, shout, cry) detected in each iteration of the duty cycle, along with a minimum threshold of foreground speech across the entire interaction recording, which resembles the real-world scenario where one may have airtime \cite{Cooney_Mastroianni_Abi-Esber_Brooks_2020} for longer period but may have the cues in most time of the interaction. These improve on previous work where the models assumed conversation turn-taking occurred at every window of 30 seconds \cite{Liang2023-gi}.
\end{itemize}

\section{Methods}
\subsection{System Development}
\noindent Based on work \cite{Zhaoyang_Sliwinski_Martire_Smyth_2018}, we define a social interaction as any instance of verbal communication between an individual and another person, whether in person, by phone, or through an online platform. To enable accurate interaction detection (Algorithm \ref{alg:socialpulse}), it is crucial to distinguish the speech of the user of the watch (foreground speech) from other audio sources, such as television or background conversations. We developed a foreground speech detector (FSD) using transfer learning with the pretrained YAMNet model \cite{noauthor_Yamnet}, and trained it on a publicly available dataset \cite{Liang_Xu_Chen_Adaimi_Harwath_Thomaz_2023} using 10-fold cross-validation. Our FSD achieved a balanced accuracy of 85. 29\% in distinguishing foreground speech from background speech, approximately 5\% higher than both the baseline in \cite{Liang_Xu_Chen_Adaimi_Harwath_Thomaz_2023} and our own previously developed model \cite{Ahmed_Rahman_Wang_Rucker_Barnes_2024}.

\begin{algorithm}
\caption{\textsc{SocialPulse}: Automatic Interaction Detection}
\label{alg:socialpulse}

\begin{algorithmic}[1]
\State $\textbf{Model}_{\!1} \gets \text{YAMNet}$,\;
         $\textbf{Model}_{\!2} \gets \text{FSD}$
\State $\mathcal C \gets$ \{\emph{Speech, Shout, Whisper, ... , Clapping}\}
\State $\Delta t \gets 1.5\ \mathrm{min}, \; L \gets 16\ \mathrm{s}$ \Comment{interval \& record length}
\State $N_{\mathrm{fs}} \gets 0$ \Comment{\# of foreground‐speech (FS)}
\State $T_{\mathrm{rec}} \gets 0$ \Comment{total recording (in seconds)}
\State $\textit{interactOn} \gets \text{false},\; \textit{startTime} \gets \bot$ 
\Statex
\While{device is worn}
    \State \textbf{wait} for $\Delta t$
    \State ${rec\_ts} \gets \textit{current\_ts}$ \Comment{ts: timestamp; rec: record}
    \State Record audio signal $a$ of length $L$
    
    \State $\text{score},\; \text{embed} \gets \textbf{Model}_{\!1}(a)$ \Comment{emebed: embeddings }
    \State $conv\_embed \gets [\;]$  \Comment{conv: conversation}
    \For{$i\gets 1\;\text{step}\;2\;\mathbf{to}\;\text{len}(\text{embed})-1$}
\State $\mathbf{\bar{s}} \gets \frac{1}{2}(\text{score}_i + \text{score}_{i+1})$ \Comment{mean of all classes}
\State ${y_i} \gets \textit{YAMNet\_classes}[\arg\max(\mathbf{\bar{s}})]$

        \If{$y_i\in\mathcal C$}
        \State $conv\_embed.\text{extend}(\mathbf embed_i, embed_{i+1})$
        \EndIf
    \EndFor
    \State $T_{\mathrm{rec}} \gets T_{\mathrm{rec}} + L$
    \State $q_{\mathrm{cue}} \gets \dfrac{\#\{y_i\in\mathcal C\}}{\lfloor \text{len}(\text{embed}) / 2 \rfloor}\times100$ \Comment{conversation cue \%}
    \If{$q_{\mathrm{cue}} < 50$}                                    \label{ln:noCue}
        \State $q_{\mathrm{fs}}  \gets \dfrac{N_{\mathrm{fs}}}{T_{\mathrm{rec}}}\times100$ \Comment{\% of FS}
        \State $\textit{cond} \gets \bigl(\textit{interactOn}\bigr)\ \wedge\ (q_{\mathrm{fs}}\ge15)$
        \If{\textit{cond} = \textbf{true} }
            \State $\textit{end} \gets \textit{rec\_ts}-\Delta t + L$ \Comment{Subtracting since no interaction was detected now.}
            \State \Call{AddInteraction}{$\textit{startTime},\textit{end}$}
        \EndIf
        \State $N_{\mathrm{fs}}\gets0,\; T_{\mathrm{rec}}\gets0, \; \textit{interactOn}\gets\textbf{false}$
    \Else  \Comment{$q_{\mathrm{cue}}\ge50$}                       \label{ln:yesCue}
        \State $class\_list \gets \textbf{Model}_{\!2}(\mathbf conv\_embed)$ \Comment{Predict FS}
         \State $N_{\mathrm{fs}} \gets N_\mathrm{fs} + \text{count 1 in class\_list} $
        
        \If{$\textit{interactOn}=\textbf{false}$}
            \State $\textit{startTime}\gets\textit{rec\_ts}, \; \textit{interactOn}\gets\textbf{true}$
        \EndIf
    \EndIf
    \EndWhile
        \If{(wearStatus=0)$\wedge$(interactOn)$\wedge$($q_{\mathrm{fs}}\ge15$)} \Comment{Watch is removed. This block is triggered based on off-body sensor's readings.}
        \State $\mathit{past} \gets \lfloor (\,current\_ts - startTime\,)/\Delta t\rfloor \times \Delta t$
        \State $end \gets startTime + past + L$
            \State \Call{AddInteraction}{$\textit{startTime},\textit{end}$}
            \State $\textit{interactOn}\gets\textbf{false}, \; N_{\mathrm{fs}}\gets0,\; T_{\mathrm{rec}}\gets0$
        \EndIf
\end{algorithmic}
\end{algorithm}

Based upon previous literature \cite{Liang2023-gi} and intuition, we identified 15 conversation cues (e.g., Chatter, Whispering, Speech, Shouting) among the 521 classes that the YAMNet model predicts. As shown in Algorithm \ref{alg:socialpulse}, our social interaction detection system triggers recording audio for 16 seconds every 1.5 minutes. The recording duration includes a 1-second padding added to the intended 15-second window to account for variability in actual recording durations, increasing the possibility of having a minimum of 15 seconds of audio data per duty cycle. The sampling strategy is designed to balance power efficiency with temporal coverage, aiming to capture as many interaction as possible while conserving energy on resource-constrained devices like smartwatches.

In contrast to existing systems such as Aware Framework \cite{noauthor_undated-aware_framework}, which records for 1 minute with 3 minutes pause, our system samples more frequently but for shorter durations. This design decision serves multiple purposes: (1) It can enhance user privacy by limiting audio capture to brief windows which can be sufficient for identifying conversational cues while minimizing exposure to sensitive content \cite{mehl2012naturalistic}; (2) It reduces energy consumption, which is critical for prolonged deployment on low-power wearable devices like a smartwatch.

Our system segments each recorded audio sample into 0.48-second frames through pre-trained YAMNet, which we used to extract acoustic embeddings and identify conversational cues. To increase resource efficiency, the FSD is not executed on all frames; instead, it is selectively applied only to a set of 2 consecutive frames at a time, which are identified as containing conversational cues. Our system labels interaction based on the percentage of foreground speech of the whole interaction (plausibly) duration (see Algorithm \ref{alg:socialpulse}) instead of relying on a specific window which may help the system to capture real-world conversation more accurately where one may interact for longer period and turn taking patterns can vary \cite{Cooney_Mastroianni_Abi-Esber_Brooks_2020}. Our system requires at least 50\% of the recorded audio to contain conversational cues during each recording cycle, which helps filter out non-conversational segments and improve detection relevance.

\subsection{Testing the System}
\noindent To test the system in the real-world environment, we developed an app incorporating the models which was deployed to Samsung Galaxy Watch 5 Pros. The system collects data in real-time and runs the models on watch. We collected data from 11 participants, comprising four undergraduate students, six graduate students, and one professor. The participants used the watches in the natural environment. The study was approved by the IRB at the University of Virginia. 

\begin{figure}
    \centering
    \includegraphics[width=1\linewidth]{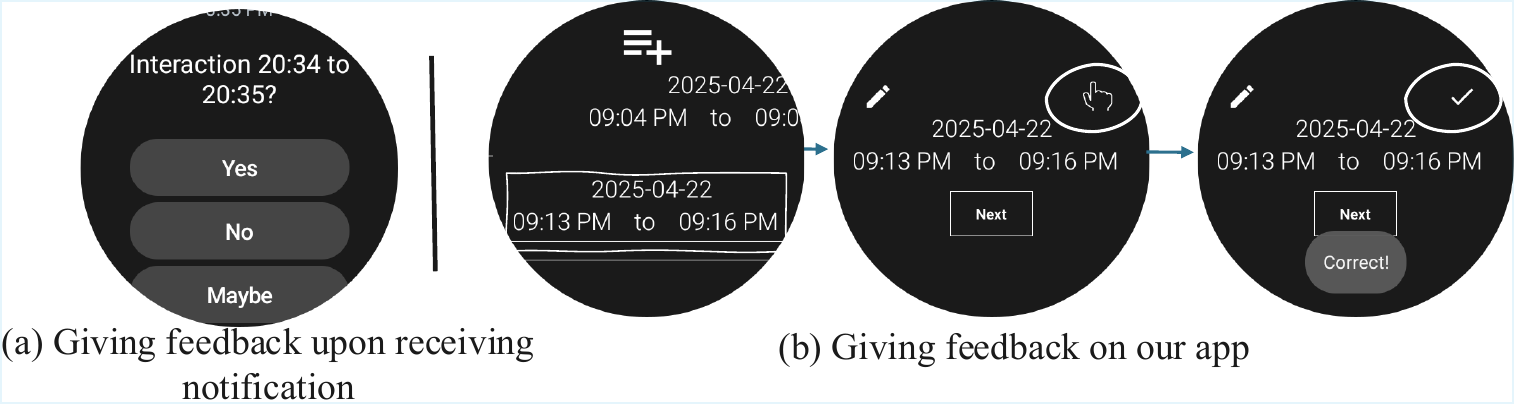}
    \caption{Options (a, b) for a participant to give feedback on the auto-detected interactions.}
    \label{fig:feedback_for_the_watch}
\end{figure}

We provided a user manual to the participants to make them aware about our system's functionality and also to inform on how to give feedback to the auto-detected interactions (Figure \ref{fig:feedback_for_the_watch}). Participants received a notification immediately after auto-detection of the end of a social interaction (Figure \ref{fig:feedback_for_the_watch}a). In cases where the notification was dismissed either by the participant or due to events such as the watch being powered off, participants could still provide feedback later via our app (Figure \ref{fig:feedback_for_the_watch}b). Feedback options included a yes or correct icon, a no or incorrect icon, and a maybe or a question-marked icon. The maybe option allowed participants to indicate uncertainty, such as when they could not accurately recall whether an interaction had occurred.

\section{Results and Discussion}
\noindent Participants experienced an average of 31.18 social interactions (SD = 17.8), ranging from 8 to 67 interactions. They used the system for 1 to 11 days, with a mean usage of 3.45 days (SD = 2.73). Over 38 participant-days, the system automatically detected 343 interactions, with participants confirming 73.18\% (N = 251) as accurate through Ecological Momentary Assessment (EMA) responses. Also, EMA data from a version deployed to five participants showed that the system correctly detected 40 virtual interactions, demonstrating it’s capability to detect both in-person and virtual interactions.

\begin{figure}
    \centering
    \includegraphics[width=1\linewidth]{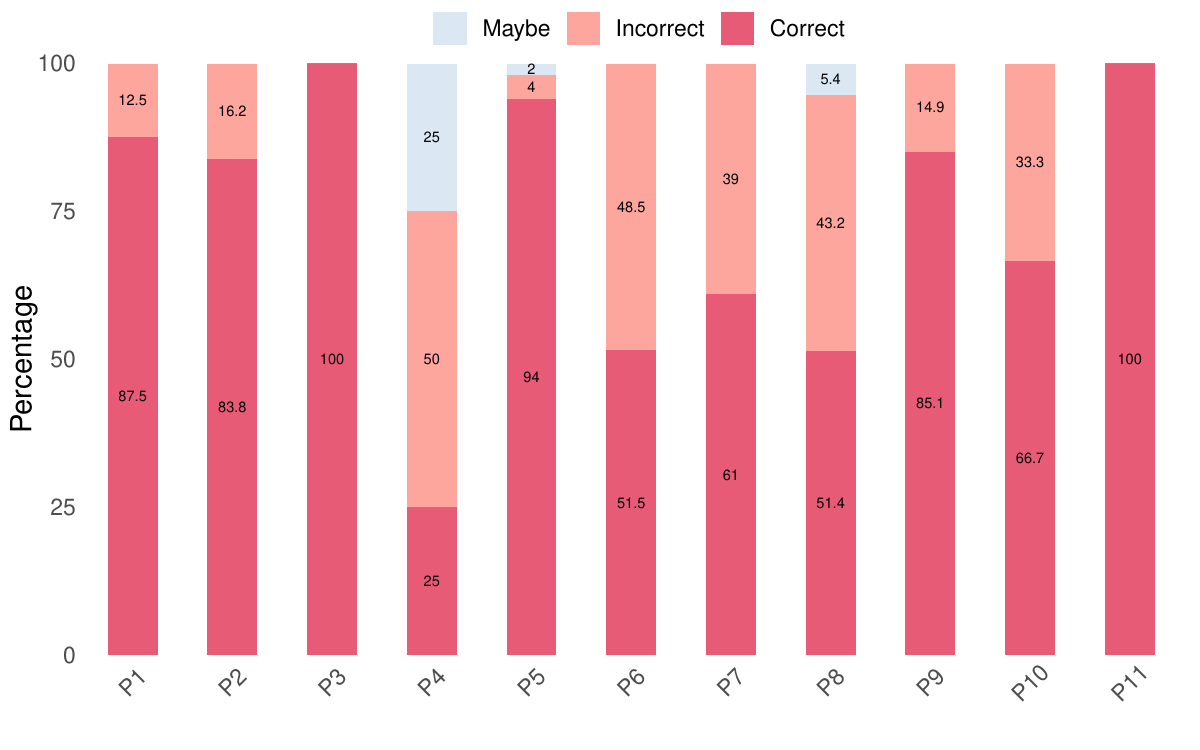}
    \caption{Performance to automatically detect interaction. P: Participant.}
    \label{fig:interaction_detect_performance}
\end{figure}

 On average, the system achieved an interaction detection accuracy per participant of 73.26\% (SD = 24.06\%), with performance ranging from 25\% to 100\% (Figure \ref{fig:interaction_detect_performance}). For participant P4, only 25\% of interactions were correctly detected, with another 25\% marked as "maybe", reflecting uncertainty. The follow-up feedback indicated that P4 frequently had a digital device playing media placed very close to the watch while at home, which likely interfered with the foreground speech detector (FSD) by making it difficult to distinguish the participant’s speech from nearby audio sources.

\begin{figure}[!b]
    \centering
    \includegraphics[width=1\linewidth]{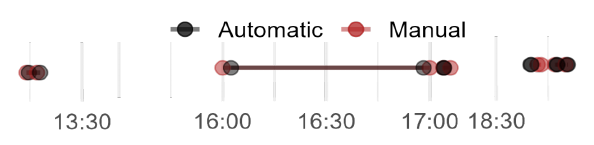}
    \caption{Difference between manually logged and auto-detected social interactions for a single participant.}
    \label{fig:compare_between_manual_and_automated}
\end{figure}

 To further evaluate the robustness of the system, we compared the initial and end times of the interaction manually logged by a participant using a notepad with those automatically detected by our system. As shown in Figure \ref{fig:compare_between_manual_and_automated}, the system demonstrated a good precision in identifying both start and end times. While the participant recorded times at the minute level, the system provided second-level precision, resulting in negligible discrepancies.

 To assess whether the system missed any interactions without triggering notifications, we followed up with six participants. All of them reported that they did not experience any interactions in which the system did not notify them. For example, a participant said, \textit{ "I did not notice any times an interaction happened where the system did not send me a notification. It worked remarkably well for me!”}. A likely reason for the absence of false negatives (that is, perfect recall) is the current configuration of the system, which triggers a notification if at least 15\% of the audio recorded during the interaction is identified as the foreground speech. This configuration of the system can be potential for real-world deployment by enabling continuous collection of labeled social interaction data with minimal participant burden, making it especially valuable for studying conditions like social anxiety.

 It should be noted that, due to the system's 1.5 minute duty cycle and the approximately 16 second audio recording window, participants were informed that the system was designed to detect interactions lasting at least 2 minutes. As a result, shorter interactions were expected to be missed. However, an analysis of 251 interactions labeled as correct revealed that 43.03\% (N = 108) were under 2 minutes and 23. 90\% (N = 60) were less than 1 minute - demonstrating the system's ability to detect even very brief interactions.

\begin{figure}
    \centering
    \includegraphics[width=0.8\linewidth]{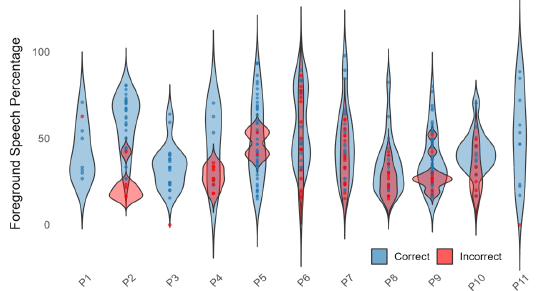}
    \caption{Foreground speech percentage in correct vs. incorrect interactions.}
    \label{fig:distribution_of_foreground_speech_percent}
\end{figure}

 As reflected in participant feedback, the system sometimes misclassified foreground speech when a nearby digital device is streaming audio. To investigate this issue, we analyzed the distribution of foreground speech percentages in detected interactions (Figure \ref{fig:distribution_of_foreground_speech_percent}). We found that for most incorrect detections, the percentage of foreground speech falls below 50\%, although this varies across interactions and participants. These findings suggest the potential of setting personalized foreground speech thresholds, instead of a fixed threshold of 15\%, which could possibly improve the accuracy of automated interaction detection.

\section{Limitations and Future Work}
\noindent While our system produced no false negatives, some participants reported issues such as segmenting longer interactions into shorter intervals. For instance, one participant noted, \textit{ "The only trouble was that it would break longer interactions into 2–5 minute intervals and ask me whether I was in an interaction many times during a single interaction."} Additionally, because the system relies solely on acoustic data, it may struggle to detect interactions when users wear headphones during virtual conversations. To enhance system performance and assess broader utility, further data collection and validation with a larger and more diverse participant pool is needed.

\section{Demonstration}
\noindent In the demo \footnote{Demo video: \href{https://github.com/BarnesLab/ACM-UbiComp-25-Demo}{https://github.com/BarnesLab/ACM-UbiComp-25-Demo}}, participants will use smartwatches equipped with our SocialPulse system. They will gain hands-on experience with real-time social interaction detection and observe the system’s capabilities in real-time.

\section*{Acknowledgments}
\noindent This work was supported in part by a 3Cavaliers Seed Grant, by the National Institute of Mental Health of the National Institutes of Health under award number R01MH132138, and the Commonwealth Cyber Initiative, an investment in the advancement of cyber R\&D, innovation, and workforce development. For more information about CCI, visit www.cyberinitiative.org.

\balance
\bibliographystyle{ACM-Reference-Format}
\bibliography{acmmobicom-ref}
\end{document}